\documentclass[aps,prl,twocolumn,superscriptaddress,showpacs]{revtex4}

\usepackage{graphicx}
\usepackage{tabularx}
\begin{document}

\title{Electron Scattering in AlGaN/GaN Structures}

\author{S. Syed}
\affiliation{Department of Applied Physics and Applied
Mathematics, Columbia University, New York, New York 10027}
\author{M. J. Manfra}
\affiliation{Bell Laboratories, Lucent Technologies, Murray Hill,
NJ 07974}
\author{Y. J. Wang}
\affiliation{National High Magnetic Field Laboratory, Florida
State University, Tallahassee, FL 32306}
\author{R. J. Molnar}
\affiliation{MIT Lincoln Laboratory, Lexington, MA 02420-0122}
\author{H. L. Stormer}\affiliation{Department of Applied Physics and Applied
Mathematics, Columbia University, New York, New York
10027}\affiliation{Bell Laboratories, Lucent Technologies, Murray
Hill, NJ 07974}\affiliation{Department of Physics, Columbia
University, New York, New York 10027}

\begin{abstract}
We present data on mobility lifetime, $\tau_t$, quantum lifetime,
$\tau_q$, and cyclotron resonance lifetime, $\tau_{CR}$, of a
sequence of high-mobility two-dimensional electron gases in the
AlGaN/GaN system, covering a density range of
$\sim1-4.5\times10^{12}$cm$^{-2}$. We observe a large discrepancy
between $\tau_q$ and $\tau_{CR}$ ($\tau_q\sim\tau_{CR}$/6) and
explain it as the result of density fluctuations of only a few
percent. Therefore, only $\tau_{CR}$ --and not $\tau_q$ -- is a
reliable measure of the time between electron scattering events in
these specimens. The ratio $\tau_t$/$\tau_{CR}$ increases with
increasing density in this series of samples, but scattering over
this density range remains predominantly in the large-angle
scattering regime.
\end{abstract}
 \maketitle

In contrast to the extensively studied two-dimensional electron
system (2DES) in AlGaAs/GaAs heterojunctions, the transport
properties of 2DES in the more recently developed AlGaN/GaN system
remain much less well understood. In particular the qualitative
nature of the scattering mechanisms at low temperatures in this
material remain controversial and the degree to which electron
scattering is preferentially large- or small-angle is still under
discussion.  Electron scattering can be characterized by an
average lifetime, $\tau$, between events and an average scattering
angle, $\phi$.  Since each scattering event dephases the
wavefunction, the quantum lifetimes, $\tau_q$, deduced from
Shubnikov de Haas (SdH) measurement and from cyclotron resonance
(CR) measurements, $\tau_{CR}$, are not expected to depend on
$\phi$, and both to be close to $\tau$. In contrast, the transport
lifetime, $\tau_t$, deduced from mobility, measures the time for
electron backscattering to occur. It depends heavily on $\phi$ and
must always exceed $\tau_q$ and $\tau_{CR}$.  For large angle
scattering, $\phi$ is large and $\tau_t$ is very similar to
$\tau_q$ and $\tau_{CR}$. If small angle scattering dominates,
then $\tau_t \gg \tau_q, \tau_{CR}$. Hence, the ratio of $\tau_t$
and $\tau_q, \tau_{CR}$ provides insight into the dominant
electron scattering events \cite{Coleridge91,DasSarmaStern85}.

A few groups have assessed the significance of one type of
scatterers over another in AlGaN/GaN heterostructures
\cite{Elhamri00,Dimitrov00,Brana00,Harris01,Frayssinet00}. Using a
sample grown by molecular beam epitaxy (MBE) on a GaN template,
Elhamri et al. \cite{Elhamri00} measured a $\tau_t/\tau_q$ ratio
of $\sim$20, suggesting the dominance of small-angle scattering in
their specimen.  Data from a single heterostructure grown on
single-crystal GaN with $n_{2D}$ = 2.4$\times10^{12}$cm$^{-2}$ and
mobility 60, 000cm$^2$/Vs indicated $\tau_t/\tau_q \sim$ 20, again
suggesting that weak scatterers play the dominant role
\cite{Frayssinet00}. A common feature of all these studies is the
reliance on SdH data to asses the inter-event lifetime, $\tau$,
which, as we will show in this report, can be unreliable in
samples with even small density inhomogeneity. Also, so far there
exist no data for $n_{2D} <$ 2$\times10^{12}$cm$^{-2}$. And even
in the mid-$10^{12}$cm$^{-2}$ range, only sparse data exist.
Lastly, there are only a few reports on CR lifetime, $\tau_{CR}$.
None of them compares the CR lifetime data to values of $\tau_t$
or $\tau_q$ of their specimens \cite{Wang96,Knap96,Knap97,Li02}.

Here we report on data for $\tau_t$, $\tau_q$, and $\tau_{CR}$
measured on heterostructures with $n_{2D}$ ranging from 1 to
4.5$\times10^{12}$cm$^{-2}$.  All measurements are performed at
$\sim$4K where, in our specimens, only scattering from static
scatterers (defects, interface roughness, residual impurities,
etc.) contribute, and scattering by phonons is negligible
\cite{phononNote}. Our results from modeling SdH oscillations
clearly indicate that $\tau_q$ is severely affected by density
inhomogeneities. We propose it not to be a good measure for the
time between scattering events in our samples. Instead,
$\tau_{CR}$ provides a good measure for this inter-event lifetime
and we can use it, in combination with $\tau_t$, to deduce the
average scattering angle, $\phi$. Our analysis shows that the
scattering events in our samples are predominantly large-angle.

\begin{table}
  \centering
  \caption{Parameters of samples discussed in this work. The density, $n_{2D}$,
  is in units of $10^{12}$ cm$^{-2}$ and the mobility, $\mu$, is in 10$^3$
  cm$^2$/Vsec}\label{table}
\newcolumntype{Y}{>{\centering\arraybackslash}X}%
\begin{tabularx}{\linewidth}{|Y|Y|Y||Y|Y|Y|}
\hline  Al$\%$ & $n_{2D}$ & $\mu$ & Al$\%$ & $n_{2D}$ & $\mu$\\
\hline 5 & 1.38 & 16 & 5 & 2.74 & 16 \\
 \hline   5 & 1.48 & 17.9 & 9 & 3.4 & 27 \\
\hline   6 & 1.6 & 12 & 10 & 3.9 & 36 \\
\hline 5 & 2.35 & 17 & 12 & 4.36& 41 \\
\hline  5 & 2.4 & 18 & & & \\
\hline
\end{tabularx}
\end{table}

Our heterostructures are grown by MBE on GaN templates prepared by
hydride vapor phase epitaxy (HVPE) on sapphire substrates.  These
templates have a typical dislocation density of
$\sim$0.5-1$\times10^9$cm$^{-2}$. The 2D density, $n_{2D}$ was
established during growth by controlling the thickness and Al$\%$
of the barrier layer. The sample parameters are listed in Table
\ref{table}. Evaporated Ti/Al contacts were used to perform van
der Pauw, low-field Hall and SdH measurements in the same
cooldown. The first two measurements were used to determine the
classical transport lifetime, $\tau_t$, from the Drude mobility
$\mu = e \tau_t / m^*$. The quantum lifetime, $\tau_q$, was
derived from SdH data using the customary expression
\cite{Coleridge91} for the oscillatory part of the
magneto-resistance. CR experiments using a Fourier transform
spectrometer were performed in a separate cooldown.  The CR
carrier lifetime, $\tau_{CR}$ is deduced from the
half-width-at-half-maximum (HWHM=$\hbar/\tau_{CR}$) of the
broadened CR line. The magnetic field was applied perpendicular to
the 2DES and the carrier density was measured \textit{in situ}
from the SdH oscillations.

\begin{figure}
\includegraphics[width=3.5in,height=2.5in]{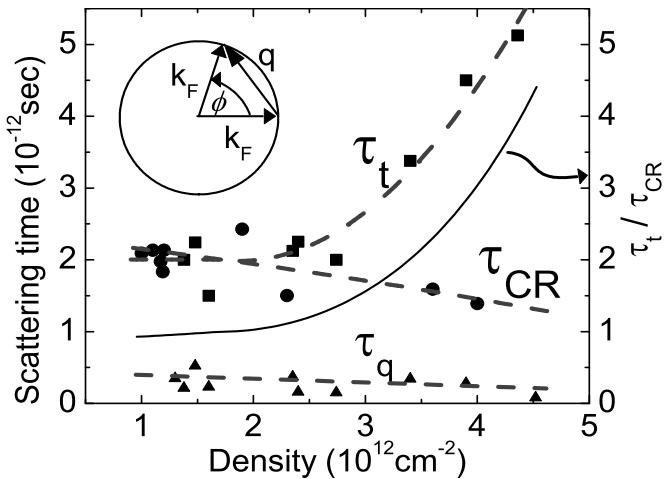}
\caption{\label{Fig.1} Experimental results of the transport,
$\tau_{t}$, quantum, $\tau_{q}$  and CR, $\tau_{CR}$, lifetimes
plotted against 2D carrier density $n_{2D}$. The dashed line
through each set of $\tau$'s is a guide to the eye. It also serves
as a parameterization of the data and is used to produce the solid
line representing the ratio $\tau_{t}/\tau_{CR}$. }
\end{figure}

Fig.\ref{Fig.1} shows the three lifetimes plotted against 2D
electron densities. In spite of the sample-to-sample scatter in
the data, there are several important observations that can be
made. In this series of samples, the mobility lifetime, $\tau_t$,
is roughly constant up to $n_{2D}\sim2\times10^{12}$cm$^{-2}$ and
then increases as the density rises to 4.5$\times10^{12}$cm$^{-2}$
The CR lifetime, $\tau_{CR}$, is very similar to $\tau_{t}$ for
$n_{2D}<$2.5$\times10^{12}$cm$^{-2}$ and then decreases slightly
for higher $n_{2D}$. The quantum lifetime, $\tau_q$, is found to
be the shortest over the entire density range studied. The dashed
lines drawn through the lifetimes are guides to the eye and are
used for parameterization. The parameterizations are required
since $\tau_{CR}$ was measured in cooldowns separate from those
for SdH and mobility measurements. This produces slightly
different densities, as seen in Fig.\ref{Fig.1} and a
parameterization of the data helps us compare the $\tau$'s.

We first address the central result of our work: the huge
discrepancy between $\tau_{CR}$ and $\tau_{q}$. Since we have
observed shifts in SdH oscillations recorded at different contacts
at the sample periphery, there is strong evidence that our
specimens are inhomogeneous. Different local densities contribute
slightly shifted oscillations in SdH and can affect the deduced
$\tau_{q}$ in a very significant way. On the other hand,
$\tau_{CR}$ is practically immune from density inhomogeneities
across the specimen. The position of the CR resonance line
$\omega_c = eB / m^*$, depends only on the electron mass, $m$*,
which is essentially density independent \cite{nonparabNote}.The
line broadening is caused almost exclusively by carrier
scattering. Hence, $\tau_{CR}$, and not $\tau_{q}$, is a good
measure of the inter-event lifetime, $\tau$.

\begin{figure}
\includegraphics[width=3.0in,height=2.5in]{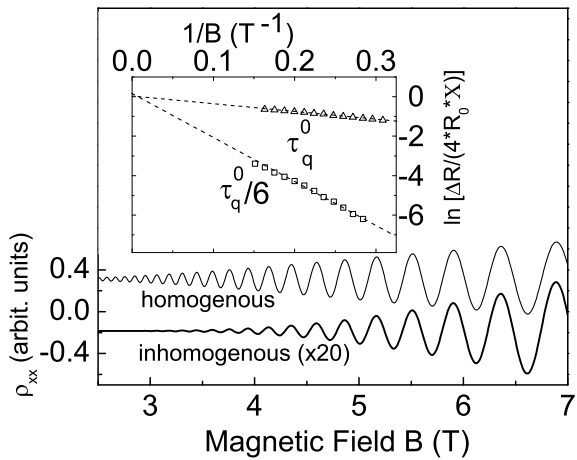}
\vspace{0.1in}
\caption{\label{Fig.2} Data simulation to model
inhomogeneity in the 2D electron density.  Magnetoresistance
calculated for uniform $n_{2D}$= 4$\times10^{12}$cm$^{-2}$
(homogeneous). The lifetime was derived from the oscillatory part
of the magneto-resistance: $\Delta R=4R_0(X/sinh X)exp
(-\pi/\omega_c\tau_q)$, with $X=2\pi^2kT/\hbar\omega_c$, $\omega_c
= eB/m^*$ the cyclotron frequency, $m$* the effective mass, and
$R_0$ the resistance at $B$=0. Convolution of these oscillations
with a Gaussian of width $\Delta n/n_{2D}$ = 2$\%$ (inhomogeneous)
yields a heavily damped set of SdH oscillations. Inset:  Quantum
lifetimes extracted from the oscillations of the homogenous and
the inhomogeneous 2DES. The 2$\%$ density inhomogeneity reduces
the original $\tau_q$ by a factor of six.}
\end{figure}

In order to model the effect of inhomogeneities we assume a
Gaussian distribution of densities of given width, $\Delta n$. The
``total SdH oscillations" are then modeled as the sum of the
distribution of all  ``partial SdH oscillations". Mathematically
this amounts to a convolution of the magnetoresistance for the
central density with a Gaussian of width $\Delta n$.  Fig.
\ref{Fig.2} shows a computer generated SdH trace for a density of
$n_{2D}$= 4$\times10^{12}$ cm$^{-2}$ and a lifetime,
$\tau_{q}^0$=1ps together with the results of a convolution with
$\Delta n$ = 0.02$n_{2D}$. Note that the convoluted trace is
heavily damped but free of ``beating''. It demonstrates that the
absence of beating in data from such high density specimens does
not guarantee a homogenous electron density. Interpreting the
convoluted trace as single-density experimental data yields the
correct average electron density of $n_{2D}$= 4$\times10^{12}$
cm$^{-2}$, but a lifetime, $\tau_{q}$= 0.18 ps, six times lower
than the actual lifetime, $\tau_{q}^0$.  Assuming $\tau_{CR}$ to
reflect the true $\tau$ we performed simulations on all our
specimens and deduce density inhomogeneities that decrease from
$\Delta n$=6$\%$ to 2$\%$ of $n_{2D}$ as the density increases
from $n_{2D}$= 1 to 4.5$\times10^{12}$ cm$^{-2}$. From our SdH
simulations, we conclude that the $\tau_{q}$'s determined from SdH
measurements are not the true quantum lifetime of the 2DES.
Instead, their relatively low value is caused by small
($\sim$2-6$\%$) density modulations in our samples. We emphasize
that the 2DES in AlGaN/GaN is particularly susceptible to this
effect due to its comparatively high electron density. In
AlGaAs/GaAs systems, typically  $n_{2D}\sim$2$\times10^{12}$
cm$^{-2}$ and the required $\Delta n$ for a similar suppression of
$\tau_{q}$ is $\sim11\%$. This is much larger than the observed
inhomogeneity ($<2\%$) in AlGaAs/GaAs structures.

For the remainder of our discussion we take $\tau_{CR}$, to be a
good measure of the inter-event lifetime, $\tau$. We compare it
with $\tau_{t}$ to learn about the average scattering angle and
wavevector, $\phi$ and $q$, respectively.  Since the 2DES is
degenerate, all scattering occurs at the Fermi energy, $E_F$,
between states of Fermi wavevector, $k_F=\sqrt{2 \pi n_{2D}}$.
Using a simple model pictured as an inset of Fig. \ref{Fig.1}, we
can write $sin (\phi/2) =  q/2k_F$. Moreover, modeling the
electron motion as a random walk, $N^2$ scatterings with average
angle $\phi=180^0/N$ are required for backscattering to occur and
hence $\tau_t\approx N^2\tau \approx(180^0/\phi)^2 \tau_{CR}$. For
$\phi=180^0$ each scattering event reverses the momentum. In this
limit, $q=2 k_F$ and $\tau_t\approx \tau_{CR}$. For
$\phi\ll180^0$, many scattering events are required for momentum
reversal. Hence, $q\ll2 k_F$  and $\tau_t\gg \tau_{CR}$. Using the
above expressions, we can deduce $\phi$ and $q$ from our data. To
this end we have parameterized the data of Fig. \ref{Fig.1} and
show their ratio, $\tau_{t}/\tau_{CR}$, in Fig. \ref{Fig.1} (see
solid line). The variation of $\tau_{t}/\tau_{CR}\approx$  1 to
4.5 reveals that the average scattering angle $\phi$ varies from
$\approx180^0$ to $\approx90^0$ as $n_{2D}$ increases from 1 to
4.5$\times10^{12}$cm$^{-2}$. It indicates that large angle
scattering dominates in the density regime of our samples. This
observation is contrary to most reports on AlGaN/GaN and is a
result of our usage of $\tau_{CR}$ rather than $\tau_q$ as a
measure for the time between scattering events. Had we used
$\tau_q$ instead, the average $\phi$ would decrease to a range of
$73^0$ to $37^0$. Using the above expressions and our
$\tau_{t}/\tau_{CR}\approx$ data, we can also derive the density
dependence of the wavevector $q$. Beyond $n_{2D}\approx$
2.5$\times10^{12}$cm$^{-2}$ where the parameterization of our data
is quite reliable (see Fig. \ref{Fig.1}) we find $q$  not to vary
significantly. This indicates that scattering events of
approximately constant $q$-value, whose average angle -- and hence
effectiveness -- decreases with increasing density, are dominant
throughout this regime.

Finally, we compare our data with available theoretical results.
From Fig.3 of the work of Hsu and Walukiewicz \cite{HsuandWalu02}
we can estimate that for $\sim$10$\%$ Al, $\tau_t$/$\tau$
increases from $\sim$10 to $\sim$20 as $n_{2D}$ increases from 1
to 2$\times10^{12}$cm$^{-2}$ and drops as the density increases
further. In contrast, we measure much smaller lifetime ratios that
rise continuously from $\sim$1 to $\sim$4.5 as $n_{2D}$ increases
from 1 to 4.5$\times10^{12}$cm$^{-2}$ (see our Fig. \ref{Fig.1}).
The discrepancies may be due to the fact that large angle
scatterers such as interface roughness and dislocations were not
considered in Ref. \cite{HsuandWalu02}. Lifetime ratios limited by
charged dislocations calculated by Jena and Mishra \cite{Jena02}
are found to be $<$10 (Ref.\cite{Jena02}, Fig.3 for
$\theta_c=\pi/10$) and monotonically increasing up to $n_{2D}\sim$
10$\times10^{12}$cm$^{-2}$. The similarities between our data and
these calculations support our model in which large angle
scattering dominates. In fact, the average spacing of dislocations
in our specimens is $\sim$400 nm as determined from AFM
measurements. The mean free path deduced from the $\tau_{CR}$ and
the Fermi velocity $v_F=\sqrt{2E_F/m^*}$ values ranges between 200
and 300 nm. This suggests dislocations to be a major contributor
to the overall low temperature scattering.

\begin{acknowledgments}
We would like to thank L.N. Pfeiffer, K.W. West, S. Das Sarma, A.
Millis and A. Mitra for helpful discussions. A portion of the work
was performed at the National High Magnetic Field Laboratory,
which is supported by NSF Cooperative Agreement No. DMR-0084173
and by the State of Florida. Financial support from the W. M. Keck
Foundation and the Office of Naval Research is gratefully
acknowledged.
\end{acknowledgments}


\begin{thebibliography}{15}
\bibitem[{non()}]{nonparabNote}
\bibinfo{note}{Even a 10$\%$ density fluctuation at 4$\times10^{12}$ cm$^{-2}$
  adds only a 1$\%$ spread in $m$* due to band non-parabolicity, which in turn
  broadens the CR line also at the 1$\%$ level.}
\end{thebibliography}

\end{document}